\documentclass[10pt]{iopart}
\usepackage{graphicx}
\usepackage{bm}
\usepackage{iopams} 
\usepackage{datetime}

\begin{document}
\title{Achieving short high-quality gate-all-around structures for horizontal nanowire field-effect transistors}

\author{J~G~Gluschke$^{1}$, J Seidl$^{1}$, A~M~Burke$^{1,2}$, R~W~Lyttleton$^{1,2}$, D~J~Carrad$^{1,3}$, A~R~Ullah$^{1,4}$, S~F~Svensson$^{2}$, S Lehmann$^{2}$, H~Linke$^{2}$ and A~P~Micolich$^{1}$}
\address{$^{1}$ School of Physics, University of New South Wales, Sydney NSW 2052, Australia}
\address{$^{2}$ Solid State Physics and NanoLund, Lund University SE-221 00 Lund, Sweden}
\address{$^{3}$ Center for Quantum Devices, Niels Bohr Institute, University of Copenhagen, Copenhagen DK-2100, Denmark}
\address{$^{4}$ School of Chemical and Physical Sciences, Victoria University of Wellington, Wellington 6021, New Zealand}
\ead{adam.micolich@nanoelectronics.physics.unsw.edu.au}

\begin{abstract}
We introduce a fabrication method for gate-all-around nanowire field-effect transistors. Single nanowires were aligned perpendicular to underlying bottom gates using a resist-trench alignment technique. Top gates were then defined aligned to the bottom gates to form gate-all-around structures. This approach overcomes significant limitations in minimal obtainable gate length and gate-length control in previous horizontal wrap-gated nanowire transistors that arise because the gate is defined by wet etching. In the method presented here gate-length control is limited by the resolution of the electron-beam-lithography process. We demonstrate the versatility of our approach by fabricating a device with an independent bottom gate, top gate, and gate-all-around structure as well as a device with three independent gate-all-around structures with 300~nm, 200~nm, and 150~nm gate length. Our method enables us to achieve subthreshold swings as low as 38~mV/dec at 77~K for a 150~nm gate length.
\end{abstract}

\submitto{\NT}
\newline
\noindent Version:~\today~~\currenttime
\maketitle
\ioptwocol

\section{Introduction}

The continuous miniaturization of the field-effect transistor (FET) has enabled the fabrication of increasingly powerful circuits on a single microchip. The performance of traditional planar FETs drops significantly as the source-drain separation is pushed below 50~nm due to short channel effects~\cite{Ferain_Nat_2011}. These short channel effects can be mitigated by improving the gate coupling~\cite{Ferain_Nat_2011, Park_IEEE_2002, Okano_IEEE_2005, Cho_CD_2004}. This led to the development of fin-FET devices with gates interfacing the channel from three sides~\cite{Ferain_Nat_2011, Park_IEEE_2002, Okano_IEEE_2005, Cho_CD_2004, Chau_NM_2007}. Optimal gate coupling is obtained from gate structures that enclose the transistor channel from all sides~\cite{Ferain_Nat_2011, Leobandung_Vac_1997, Colinge_IEEE_1990}. However, these structures can be difficult to make using the conventional top-down fabrication techniques employed for planar devices.~\cite{Leobandung_Vac_1997, Colinge_IEEE_1990, Singh_IEEE_2006}. 

A bottom-up approach exploiting self-assembled nanowires offers a simpler pathway to fully conformal `gate-all-around' structures~\cite{Samuelson_MT_2003}. These nanowires stand vertically, enabling a conformal coating of gate oxide and gate metal to be applied in a straightforward way~\cite{Tanaka_APE_2010, Bryllert_IEEE_2006, Ng_NL_2004,Storm_NL_2011, Burke_NL_2015}. The nanowires can be processed into vertical FET arrays on the growth substrate~\cite{Tanaka_APE_2010, Bryllert_IEEE_2006, Ng_NL_2004} or can be transferred to a separate device substrate to create horizontal devices~\cite{Storm_NL_2011, Burke_NL_2015}. Vertical nanowire arrays have achieved near thermal-limit subthreshold swings~\cite{Bryllert_DRCD_2005}, integration of III-Vs on Si~\cite{Tomioka_Nat_2012}, and continue down a road towards practical applications~\cite{Riel_MRS_2014}. This orientation has also seen work to incorporate heterostructured nanowires towards high-performance tunnel-FETs~\cite{Memisevic_IEEE_2017, Memisevic_IEEE_2018} and produce multiple independent 'wrap gates'~\cite{Li_IEEE_2011}. Turning to horizontal wrap-gated nanowire transistors, these are of interest for basic research devices, e.g., quantum electronics, but also as a possible complement for vertical transistors in 3D-integrated circuits~\cite{Ferry_Sci_2008}. Fabrication of multiple independent wrap-gates has also been achieved in the horizontal orientation, giving a significant advantage on scalability over the vertical orientation~\cite{Burke_NL_2015}.

A major limitation of horizontal wrap-gate nanowire transistors~\cite{Storm_NL_2011, Burke_NL_2015} is that the gate length is defined by wet-etching. This limits control and restricts the minimum achievable gate length~\cite{Burke_NL_2015}. Shorter gates are also intrinsically of lower quality in this instance because unintentional overetching introduces `mouse-bite' defects---small holes in the gate metal and oxide that compromise both performance and yield~\cite{Burke_NL_2015}. Burke \textit{et al.}~\cite{Burke_NL_2015} found that the shortest gate length that could be reliably achieved was $\sim$300~nm. The minimal gate length is important for electronics applications and basic research. For industrial applications, the gate length governs the density of devices on a microchip. For research studies, sub-200~nm gates are desirable for nanowire quantum devices e.g. gate-defined quantum dots~\cite{Pfund_APL_2006, Fasth_NL_2007} and nanowire quantum-point contacts~\cite{Abay_NL_2013, Heedt_NL_2016a, Heedt_NL_2016b, Saldana_NL_2018}.

\begin{figure*}
\centering
\includegraphics[width=1\textwidth]{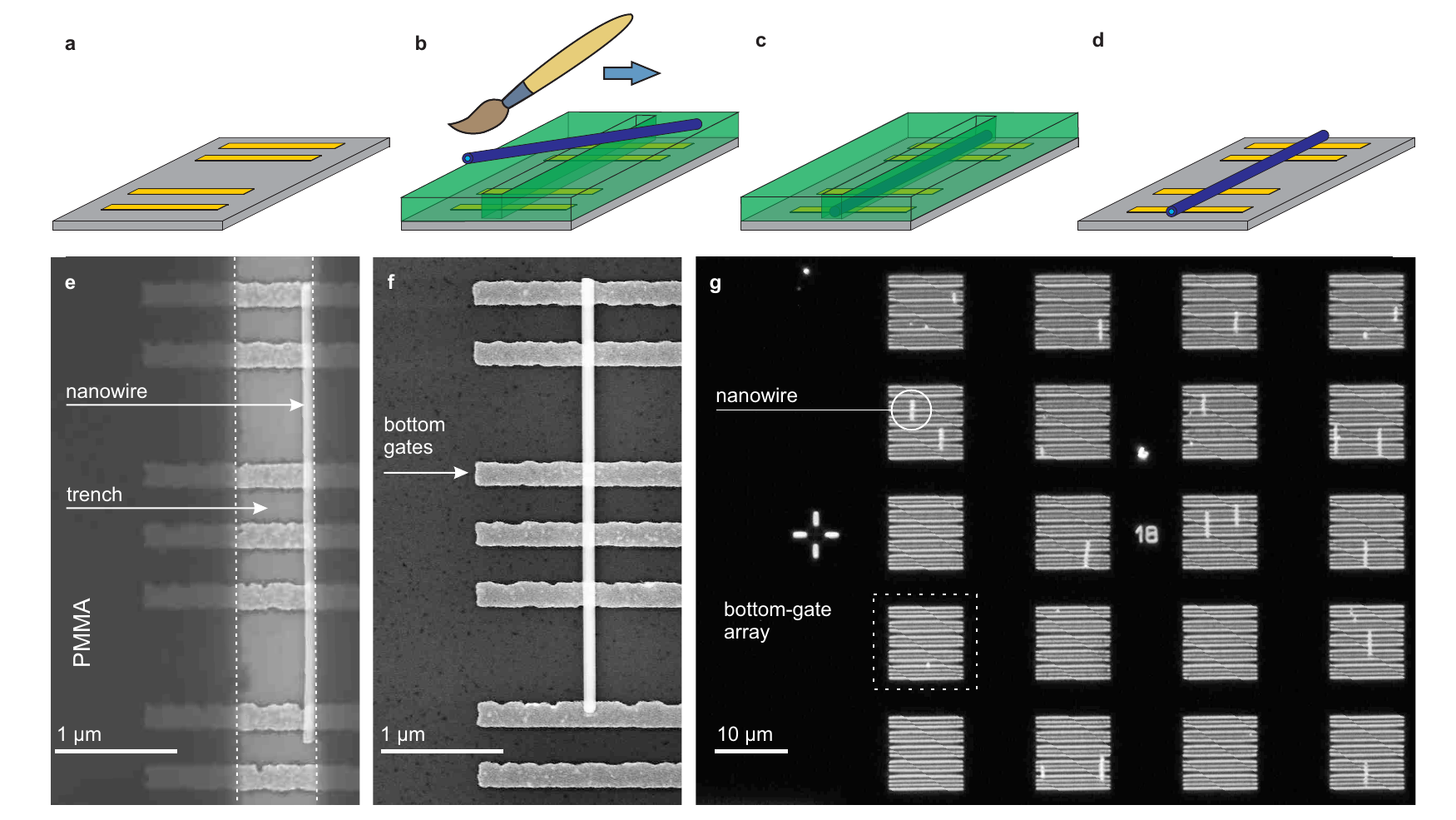}
\caption{Illustration of the nanowire alignment procedure. (a) Arrays of 30~nm thick bottom gates were patterned by electron-beam lithography (EBL). (b) and (c) 300~nm wide trenches were defined in $\sim$300~nm thick EBL resist. Nanowires were deposited on top of the resist, covered with a drop of isopropyl alcohol and then brushed into the trenches~\cite{Lim_Small_2010, Lard_NL_2014}. (d) The nanowires remain aligned perpendicular to the bottom gates after removing the resist with acetone. Scanning-electron microscopy image (e) of a nanowire inside a resist trench and (f) a different nanowire after the resist is removed. (g) Dark-field optical microscopy image of nanowires aligned to bottom gates. \label{fig:F1_align}}
\end{figure*}

Here, we introduce a fabrication process for horizontal nanowire FETs with multiple gate-all-around structures that maintain the advantages of the horizontal orientation while overcoming the limitations arising from wet etching. The gate length is defined instead by electron-beam lithography (EBL) patterned metal deposition. This presents the challenge of obtaining the portion of the gate directly underneath the nanowire. We achieve this by depositing nanowires on pre-fabricated bottom gates before completing the gate-all-around structure by depositing a top gate aligned with the bottom gate. A crucial aspect of the fabrication process is that the nanowire is aligned perpendicular to the bottom gate. This alignment is achieved with high accuracy using a resist-trench technique (see figure~1)~\cite{Lim_Small_2010, Lard_NL_2014}. As a result, the minimal achievable gate length, control over gate length, and gate-metal quality are limited by the EBL process rather than wet-etch steps. At this point we note a nomenclature distinction regarding wrap-gates and gate-all-around structures. The wrap-gate devices~\cite{Storm_NL_2011, Burke_NL_2015} have unambiguously conformal gate metallisation. In contrast, our process has the possibility of small voids under the nanowire edges (see figure~2(e)). For clarity, in distinguishing between them, we refer to our devices as gate-all-around structures rather than wrap gates.

We demonstrate the full capacity of our method with two devices. The first is a single nanowire with independently controllable bottom gate, top gate, and gate-all-around structures. It highlights that the strongest gating is obtained with a gate-all-around structure, but that an $\Omega$-shaped top gate gives comparable performance in a side-by-side comparison. This is consistent with modelling predictions~\cite{Tang_IEEE_2004, Li_IEEE_2005}. The second device is a single nanowire with three independently controllable gate-all-around structures with different gate lengths: 300~nm, 200~nm, and 150~nm. This is a significant improvement in minimal gate length compared to earlier horizontal wrap gates~\cite{Storm_NL_2011, Burke_NL_2015}.

\section{Experimental details}

\subsection{Nanowire alignment}

Our devices are made on an n$^{+}$-Si substrate capped with 100~nm of SiO$_{2}$ and 10~nm of HfO$_{2}$. The HfO$_{2}$ layer serves as an etch-stop layer when the Al$_{2}$O$_{3}$ gate insulator is etched during contact formation. Arrays of bottom gates were patterned by EBL and evaporation of Ni/Au (5/25~nm), as shown in figure~1(a). The bottom gates are 150-300~nm wide, 10~$\mu$m long and have variable inter-gate spacing. There are 20 bottom-gate arrays per 100$\times$100~$\mu$m$^{2}$ device field (figure~1(g)) to enable satisfactory device yield.

The nanowires are positioned using a resist-trench method~\cite{Lim_Small_2010, Lard_NL_2014}, as follows: The substrate was spin-coated with $\sim$300~nm of \textit{MicroChem} polymethyl methacrylate (PMMA) 950k A5 EBL resist. Trenches with length 10~$\mu$m and width 200-400~nm were defined by EBL. These trenches are perpendicular to the underlying bottom gates (see figure~1(e)). Any resist residue in the trenches was removed with a 30~s oxygen-plasma etch after development (50~W, 340~mTorr). Wurtzite InAs nanowires approximately 50~nm in diameter and 3 to 10~$\mu$m long were grown by chemical beam epitaxy (CBE)~\cite{Jensen_NL_2004} or metal organic vapour phase epitaxy (MOVPE)~\cite{Lehmann_NL_2013}. They were conformally coated with a 16~nm Al$_{2}$O$_{3}$ gate dielectric by atomic layer deposition (ALD). The oxide coating of the nanowire removes the need to cover the bottom gates with an insulator as done previously~\cite{Fasth_NL_2007, Saldana_NL_2018}. The nanowires were picked up from the growth substrate with the tip of a triangular piece of clean-room tissue and deposited on top of the patterned resist. Approximately 20-50 nanowires were transferred to each of the 24 100$\times$100~$\mu$m$^2$ regions with 20 bottom-gate arrays per 100$\times$100~$\mu$m$^2$ region (figure~\ref{fig:F1_align}(g)). The substrate was then covered in a single droplet of isopropyl alcohol. A piece of clean-room tissue was used to brush the nanowires into the trenches until the isopropyl alcohol evaporated completely (see figure~\ref{fig:F1_align}(b) and (c)). This process was repeated 2-4 times until no nanowires were visible on top of the resist near the area patterned with trenches under a dark-field microscope. Finally the PMMA resist was removed in an acetone bath leaving the aligned nanowires adhered to the bottom-gate array (see figure~\ref{fig:F1_align}(d)). Any nanowires left on top of the resist were washed away, leaving only aligned nanowires. Empty trenches cannot be distinguished from trenches with nanowires using a 1000$\times$ optical microscope prior to the removal of the resist. 

We estimated the yield of the alignment procedure by counting the nanowires in seven 100$\times$100~$\mu$m$^{2}$ regions each with 100 trenches after the initial deposition and again after removal of the resist. Typically, 5-30\% of 20-50 distributed nanowires were aligned successfully (see figure~\ref{fig:F1_align}(g)). This yield is sufficient as a complete device only requires one nanowire per 100$\times$100~$\mu$m$^{2}$ region. Lard \textit{et al.}~\cite{Lard_NL_2014} have demonstrated the assembly of highly-ordered nanowire arrays with this method using higher nanowire density and fine-tuning of the trench dimensions. This demonstrates the scalability of this approach, which could be used, e.g., to prototype integrated nanowire circuits with multiple nanowires on the same chip.

In this study, the trench width had no significant impact on the yield of captured nanowires for trench widths of 200~nm, 300~nm, and 400~nm. We rarely observed multiple nanowires captured in the same trench. This is likely due to the large supply of trenches relative to the number of available nanowires. An unexpected finding was that the accuracy of angular nanowire alignment was independent of trench width. We attribute this to the nanowires sticking to the trench side-walls during capture, resulting in optimal angular alignment for nearly all nanowires (see figure~\ref{fig:F1_align}(e)). The orientation is generally maintained upon removal of the resist as shown in figure~\ref{fig:F1_align}(f).

\subsection{Nanowire contacts} \label{sub:Contacts}

\begin{figure}
\centering
\includegraphics[width=1\columnwidth]{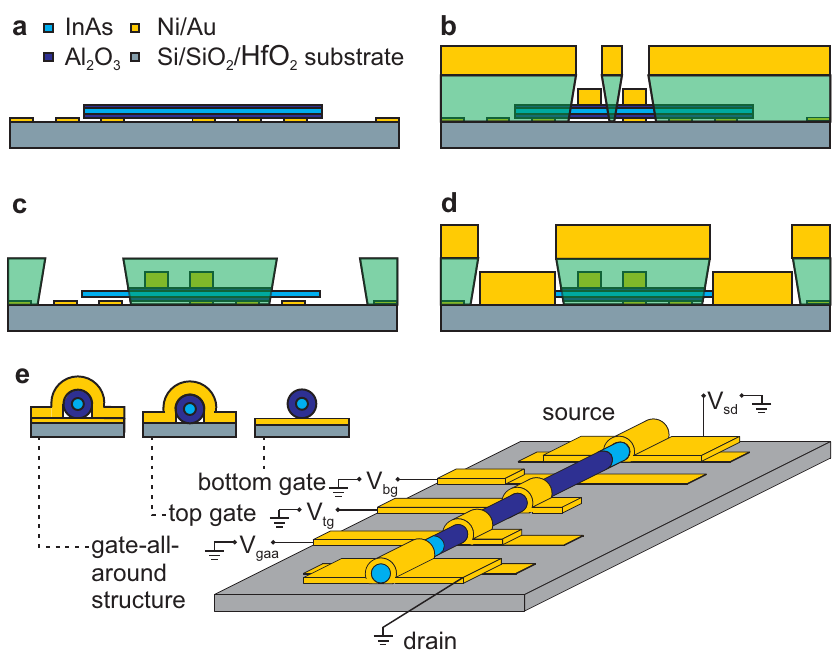}
\caption{Schematic of the fabrication of a nanowire field-effect transistors with three different gate types. (a) An Al$_{2}$O$_{3}$ coated nanowire was placed orthogonally on an array of bottom gates (figure~\ref{fig:F1_align}). (b) Top gates were defined using electron-beam lithography (EBL) and metal evaporation. A gate-all-around structure was formed where top and bottom gates are aligned. (c) The source and drain contacts were exposed in EBL resist and the Al$_{2}$O$_{3}$ coating was removed with a HF etch. (d) The source and drain contacts were passivated with (NH$_{4}$)$_{2}$S$_{x}$ immediately prior to metallization. (e) The finished device has an independent gate-all-around structure, top gate and bottom gate on the same nanowire. \label{fig:F2_fab}}
\end{figure}

The fabrication process for the source, drain, and gate electrodes for a device with a gate-all-around structure, a top gate, and a bottom gate is shown in figure~\ref{fig:F2_fab}. Figure~\ref{fig:F3_3gate}(a) shows the finished device. The same processing steps can be applied to create devices with multiple gate-all-around structures such as the nanowire FET shown in figure~\ref{fig:F4_length}(a). The substrate with the aligned nanowires (figure~\ref{fig:F2_fab}(a)) was once more coated with EBL resist. Top gates were exposed and metallized (Ni/Au, 6/134 nm) after a 30~s oxygen-plasma treatment to remove any resist residue (see figure~\ref{fig:F2_fab}(b)). Excess metal was removed together with the EBL resist in an acetone lift-off at 60$^{\circ}$C. Source and drain contacts were exposed in a final EBL step. 
The Al$_{2}$O$_{3}$ coating was removed from the exposed nanowire ends by a 15~s buffered HF etch (1:7 HF:NH$_{4}$F) as shown in figure~\ref{fig:F2_fab}(c). Wet-etching can be eliminated by substituting the gate oxide with the organic gate insulator parylene, which can be removed by oxygen-plasma etching~\cite{Gluschke_NL_2018}. The sample was treated with (NH$_{4}$)$_{2}$S$_{x}$ solution immediately prior to the metal deposition by thermal evaporation (Ni/Au, 6/134 nm) to ensure ohmic contacts~\cite{Suyatin_NT_2007}. Excess metal was removed in an acetone lift-off at 60~$^{\circ}$C giving the completed device shown in figure~\ref{fig:F2_fab}(e).

\subsection{Electrical measurements}

All electrical measurements were performed in liquid nitrogen (temperature $T$~=~77~K) to improve gate stability and reduce hysteresis due to charge trapping at the Al$_{2}$O$_{3}$-InAs interface~\cite{Burke_NL_2015}. A dc source-drain voltage $V_{sd}$~=~50~mV was applied at the source contact to drive a drain current $I_{d}$ measured using a \textit{Keithley} 6517A electrometer at the drain. The gate voltage $V_{g}$ was applied using the dc auxiliary ports of a \textit{Stanford Research} SR830 lock-in amplifier after confirming negligible gate leakage ($<100$~pA) for the individual gates with a \textit{Keithley} K2401 source-measure unit. $I_{d}$ was recorded for decreasing $V_{g}$. Only one gate was swept at a time with all other gates kept grounded.

\section{Results and discussion}

\subsection{Bottom, top, and gate-all-around structure}

\begin{figure}
\centering
\includegraphics[width=1\columnwidth]{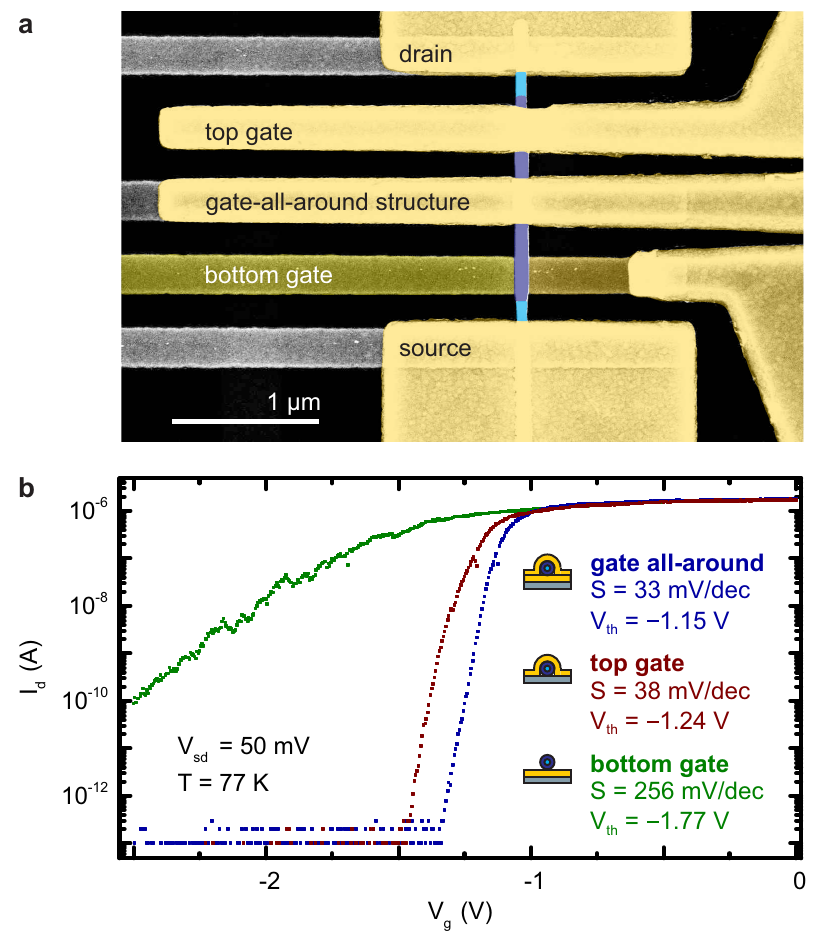}
\caption{(a) False-coloured scanning-electron microscopy image of a device with three different gate types: a top gate, a gate-all-around structure and a bottom gate. (b) Drain current $I_{d}$ vs gate voltage $V_{g}$ for the three different gate types on the same nanowire. \label{fig:F3_3gate}}
\end{figure}

Figure~\ref{fig:F3_3gate}(a) shows an scanning-electron microscopy image of a device with three different gate types: a top gate, a gate-all-around structure, and a bottom gate. All three gates are approximately 250~nm in length. The EBL process yielded smooth, conformal metal gates without the `mouse-bite' defects found in short horizontal wrap gates~\cite{Burke_NL_2015}. Overlapping top and bottom gates are aligned to within 10~nm. The top gates are up to 20~nm wider than the bottom gates because top gate evaporation was carried out at an angle of 15-20$^{\circ}$ under rotation to ensure gate continuity across the nanowire. If required, this can be compensated for by reducing the width of the top gates in the EBL pattern. We estimate the coverages as 100\% for the gate-all-around structure, 73\% for the top gate, and 17\% for the bottom gate based on geometrical considerations.

Figure~\ref{fig:F3_3gate}(b) shows the electrical performance of the three individual gates from a nominally identical device. The gate-all-around structure gives the steepest subthreshold swing $S$~=~33~mV/dec. This is approximately twice the thermal limit of 15.3~mV/dec at 77~K and competitive with the $S$~=~25 to 43~mV/dec reported by Burke \textit{et al.}~\cite{Burke_NL_2015} for InAs nanowire FETs with longer, etch-defined wrap-gates at 77~K. The $\Omega$-shaped top gate performs almost as well giving $S$~=~38~mV/dec. This is consistent with modelling predictions~\cite{Tang_IEEE_2004, Li_IEEE_2005}. The bottom gate performs significantly worse with $S$~=~256~mV/dec due to reduced gate coupling resulting from the limited gate coverage of the nanowire circumference. The reduced gate coupling means that higher gate voltages are required to deplete the nanowire. This causes a shift in threshold voltage $V_{th}$ to more negative values. The shift is small for the top gate but larger for the bottom gate.

\subsection{Different gate lengths}

\begin{figure}
\centering
\includegraphics[width=1\columnwidth]{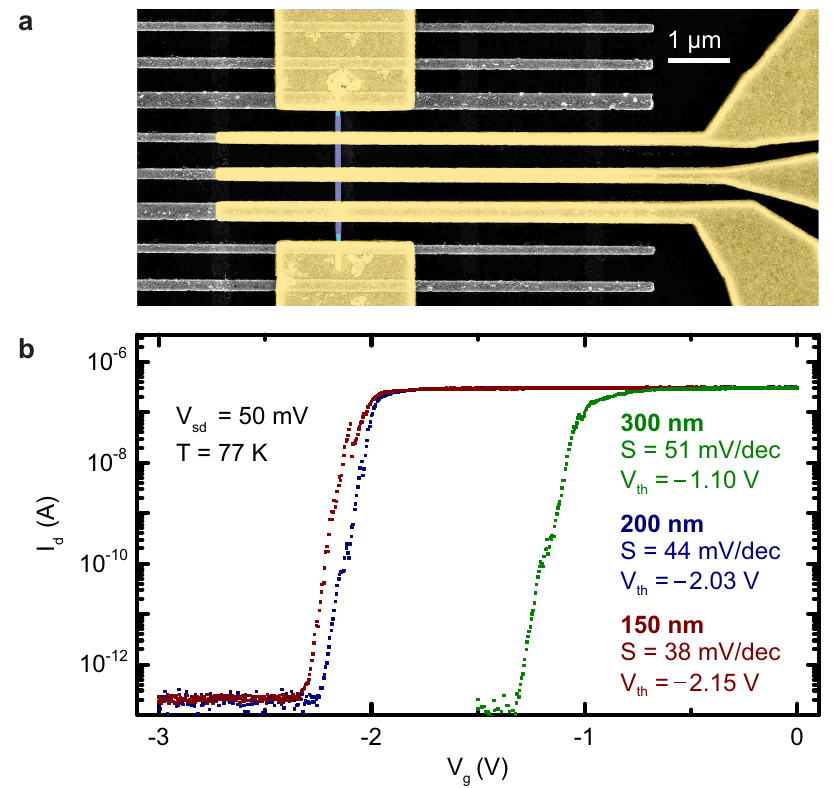}
\caption{(a) False-coloured scanning-electron microscopy image of a device with three gate-all-around structures with different gate lengths. (b) Drain current $I_{d}$ vs gate-all-around voltage $V_{g}$ for a nominally identical device.\label{fig:F4_length}}
\end{figure}

A device with three different length gate-all-around structures demonstrates the enhanced control over gate length and ability to make shorter gates. Figure~\ref{fig:F4_length}(a) shows a false-coloured SEM image of a device with three gate-all-around structures 300~nm, 200~nm, and 150~nm long. Data from a nominally identical device is displayed in figure~\ref{fig:F4_length}(b). The threshold voltage $V_{th}$~=~$-1.10$~V for the 300~nm gate-all-around structure is similar to the $V_{th}$~=~$-1.15$~V obtained for the 250~nm gate-all-around structure in figure~\ref{fig:F3_3gate}. $V_{th}$ is shifted significantly to larger negative $V_{th}$ for the 150~nm and 200~nm gates indicating decreased effective gate coupling per gate length. Interestingly, no degradation in subthreshold swing is observed with decreased gate length. In fact, $S$ improves slightly for the shorter gates with $S$~=~52~mV/dec for the 300~nm long gate, 45~mV/dec for the 200 nm gate, and 37~mV/dec for the 150~nm gate. A similar behaviour was observed by Burke \textit{et al.}~\cite{Burke_NL_2015} in wrap-gated InAs nanowire transistors at 77~K (see supplementary information). Three-dimensional electrostatic-modelling studies have shown that effective gate capacitance per length decreases for shorter gates due to fringing effects~\cite{Ji_ME_2008, Song_IEEE_2006, Zou_IEEE_2011, Gupta_JS_2013}. This leads to a shift in threshold voltage to more negative values as the gate length decreases. This shift is non-linear and increases with reduced length~\cite{Park_IEEE_2002, Ji_ME_2008, Song_IEEE_2006}. Generally, drain-induced barrier lowering also contributes, driving an associated degradation in subthreshold swing~\cite{Park_IEEE_2002, Song_IEEE_2006}. This is clearly not observed in figure~4(b) and we speculate that the small source-drain bias and the strong gate coupling~\cite{Ferain_Nat_2011, Park_IEEE_2002, Leobandung_Vac_1997, Lee_SSE_2007} make this effect insignificant for the gate lengths studied here. This effect may become significant for gate lengths $<$100~nm. Electrostatic simulations~\cite{Heedt_NS_2015} for this aspect of these devices could generate further insight and are encouraged.

\section{Conclusion}

We introduced a versatile fabrication technique for gate-all-around structure nanowire FETs. Single nanowires were positioned perpendicularly on top of pre-defined bottom gates using a resist-trench alignment technique~\cite{Lim_Small_2010, Lard_NL_2014}. Top gates were then created in alignment with bottom gates to form gate-all-around structures. This approach overcomes a key limitation of established wrap-gate methods~\cite{Storm_NL_2011, Burke_NL_2015} where a metal etch is used to define gate segments; namely the limitation of gate-length control and minimal gate length due to over-etching. Gate length and quality in our approach are only limited by the resolution of the EBL process. We demonstrated the length control by fabricating a device with independent 300~nm, 200~nm, and 150~nm long gate-all-around structures with a subthreshold swing of 38~mV/dec at 77~K for the 150~nm gate. We expect process optimization will yield further significant reduction in minimal gate length as sub-20~nm features can be achieved with commercial EBL systems \cite{Cord_NSPMP_2009,Mohammad_book_2010}. This platform may be interesting to systematically study gate-length dependent transistor performance. Our process also enables the fabrication of multiple gate types such as gate-all-around structures, top gates and bottom gates on the same nanowire. The gate-all-around structure performed best followed by the top gate and then the bottom gate. This is expected due to the different electrostatic couplings of the gate geometries~\cite{Tang_IEEE_2004, Heedt_NS_2015}. 

\ack This work was funded by the Australian Research Council (ARC) under DP170102552 and DP170104024, UNSW Goldstar Scheme, NanoLund at Lund University, Swedish Research Council, Swedish Energy Agency (Grant No. 38331-1) and Knut and Alice Wallenberg Foundation (KAW). APM acknowledges an ARC Future Fellowship (FT0990285). This work was performed in part using the NSW node of the Australian National Fabrication Facility (ANFF).


\section*{References}

\begin{thebibliography}:

\bibitem{Ferain_Nat_2011} Ferain I, Colinge C A and Colinge J-P 2011 Multigate transistors as the future of classical metal-oxide-semiconductor field-effect transistors \textit{Nature} \textbf{479} 310-6

\bibitem{Park_IEEE_2002} Park J-T and Colinge J-P 2002 Multiple-gate SOI MOSFETs: Device design guidelines \textit{IEEE Trans. Electron Devices} \textbf{49} 2222-9

\bibitem{Okano_IEEE_2005} Okano K \textit{et al.} 2005 Process integration technology and device characteristics of CMOS FinFET on bulk silicon substrate with sub-10 nm fin width and 20 nm gate length \textit{IEEE International Electron Devices Meeting, 2005.} pp 721-4

\bibitem{Cho_CD_2004} Cho H J, Choe J D, Li M, Kim J Y, Chung S H, Oh C W, Yoon E-J, Kim D-W, Park D and Kim K 2004 Fin width scaling criteria of body-tied FinFET in sub-50 nm regime \textit{62nd DRC. Conference Digest Device Research Conference, 2004.} pp 209-10 vol.1

\bibitem{Chau_NM_2007} Chau R, Doyle B, Datta S, Kavalieros J and Zhang K 2007 Integrated nanoelectronics for the future \textit{Nat. Mater.} \textbf{6} 810-2

\bibitem{Leobandung_Vac_1997} Leobandung E 1997 Wire-channel and wrap-around-gate metal-oxide-semiconductor field-effect transistors with a significant reduction of short channel effects \textit{J. Vac. Sci. Technol. B} \textbf{15} 2791

\bibitem{Colinge_IEEE_1990} Colinge J-P, Gao M H, Romano-Rodriguez A, Maes H and Claeys C 1990 Silicon-on-insulator “gate-all-around device” \textit{International Technical Digest on Electron Devices} pp 595-8

\bibitem{Singh_IEEE_2006} Singh N, Agarwal A, Bera L K, Liow T Y, Yang R, Rustagi S C, Tung C H, Kumar R, Lo G Q, Balasubramanian N and Kwong D-L 2006 High-performance fully depleted silicon nanowire (diameter $\leq$ 5 nm) gate-all-around CMOS devices \textit{IEEE Electron Device Lett.} \textbf{27} 383-6

\bibitem{Samuelson_MT_2003} Samuelson L 2003 Self-forming nanoscale devices \textit{Mater. Today} \textbf{6} 22-31

\bibitem{Tanaka_APE_2010} Tanaka T, Tomioka K, Hara S, Motohisa J, Sano E and Fukui T 2010 Vertical surrounding gate transistors using single InAs nanowires grown on Si substrates \textit{Appl. Phys. Express} \textbf{3} 025003

\bibitem{Bryllert_IEEE_2006} Bryllert T, Wernersson L E, Fr\"oberg L E and Samuelson L 2006 Vertical high-mobility wrap-gated InAs nanowire transistor \textit{IEEE Electron Device Lett.} \textbf{27} 323-5

\bibitem{Ng_NL_2004} Ng H T, Han J, Yamada T, Nguyen P, Chen Y P and Meyyappan M 2004 Single crystal nanowire vertical surround-gate field-effect transistor \textit{Nano Lett.} \textbf{4} 1247-52

\bibitem{Storm_NL_2011} Storm K, Nylund G, Samuelson L and Micolich A P 2011 Realizing lateral wrap-gated nanowire FETs: Controlling gate length with chemistry rather than lithography \textit{Nano Lett.} \textbf{12} 1-6

\bibitem{Burke_NL_2015} Burke A M, Carrad D J, Gluschke J G, Storm K, Fahlvik-Svensson S, Linke H, Samuelson L and Micolich A P 2015 InAs nanowire transistors with multiple, independent wrap-gate segments \textit{Nano Lett.} \textbf{15} 2836-43

\bibitem{Bryllert_DRCD_2005} Bryllert T, Samuelson L, Jensen L E and Wernersson L E 2005 Vertical high mobility wrap-gated InAs nanowire transistor \textit{Device Research Conference Digest} \textbf{1} pp 157-8

\bibitem{Tomioka_Nat_2012} Tomioka K, Yoshimura M and Fukui T 2012 A III–V nanowire channel on silicon for high-performance vertical transistors \textit{Nature} \textbf{488} 189-92

\bibitem{Riel_MRS_2014} Riel H, Wernersson L-E, Hong M and del Alamo J A 2014 III-V compound semiconductor transistors---from planar to nanowire structures \textit{MRS Bull.} \textbf{39} 668-77

\bibitem{Memisevic_IEEE_2017} Memisevic E, Svensson J, Lind E and Wernersson L 2017 InAs/InGaAsSb/GaSb nanowire tunnel field-effect transistors \textit{IEEE Trans. Electron Devices} \textbf{64} 4746-51

\bibitem{Memisevic_IEEE_2018} Memisevic E, Svensson J, Lind E and Wernersson L 2018 Vertical nanowire TFETs with channel diameter down to 10 nm and point S$_{MIN}$ of 35 mV/decade \textit{IEEE Electron Device Lett.} \textbf{39} 1089-91

\bibitem{Li_IEEE_2011} Li X, Chen Z, Shen N, Sarkar D, Singh N, Banerjee K, Lo G-Q and Kwong D-L 2011 Vertically stacked and independently controlled twin-gate MOSFETs on a single Si nanowire \textit{IEEE Electron Device Lett.} \textbf{32} 1492-4

\bibitem{Ferry_Sci_2008} Ferry D K 2008 Nanowires in nanoelectronics \textit{Science} \textbf{319} 579-80

\bibitem{Pfund_APL_2006} Pfund A, Shorubalko I, Leturcq R and Ensslin K 2006 Top-gate defined double quantum dots in InAs nanowires \textit{Appl. Phys. Lett.} \textbf{89} 252106

\bibitem{Fasth_NL_2007} Fasth C, Fuhrer A, Samuelson L, Golovach V N and Loss D 2007 Direct measurement of the spin-orbit interaction in a two-electron InAs nanowire quantum dot \textit{Phys. Rev. Lett.} \textbf{98} 266801

\bibitem{Abay_NL_2013} Abay S, Persson D, Nilsson H, Xu H Q, Fogelstr\"om M, Shumeiko V and Delsing P 2013 Quantized conductance and its correlation to the supercurrent in a nanowire connected to superconductors \textit{Nano Lett.} \textbf{13} 3614-7

\bibitem{Heedt_NL_2016a} Heedt S, Prost W, Schubert J, Gr\"utzmacher D and Sch\"apers T 2016 Ballistic transport and exchange interaction in InAs nanowire quantum point contacts \textit{Nano Lett.} \textbf{16} 3116-23

\bibitem{Heedt_NL_2016b} Heedt S, Manolescu A, Nemnes G A, Prost W, Schubert J, Gr\"utzmacher D and Sch\"apers T 2016 Adiabatic edge channel transport in a nanowire quantum point contact register \textit{Nano Lett.} \textbf{16} 4569-75

\bibitem{Saldana_NL_2018} Estrada Salda\~na J C, Niquet Y-M, Cleuziou J-P, Lee E J H, Car D, Plissard S R, Bakkers E P A M and De Franceschi S 2018 Split-channel ballistic transport in an InSb nanowire \textit{Nano Lett.} \textbf{18} 2282-7

\bibitem{Lim_Small_2010} Lim J K, Lee B Y, Pedano M L, Senesi A J, Jang J-W, Shim W, Hong S and Mirkin C A 2010 Alignment strategies for the assembly of nanowires with submicron diameters \textit{Small} \textbf{6} 1736-40

\bibitem{Lard_NL_2014} Lard M, ten Siethoff L, Generosi J, M\aa nsson A and Linke H 2014 Molecular motor transport through hollow nanowires \textit{Nano Lett.} \textbf{14} 3041-6

\bibitem{Tang_IEEE_2004} Tang C-S, Yu S-M, Chou H-M, Lee J-W and Li Y 2004 Simulation of electrical characteristics of surrounding- and omega-shaped-gate nanowire FinFETs \textit{4th IEEE Conference on Nanotechnology}, 2004. pp 281–3

\bibitem{Li_IEEE_2005} Li Y, Chou H-M and Lee J-W 2005 Investigation of electrical characteristics on surrounding-gate and omega-shaped-gate nanowire FinFETs \textit{IEEE Trans. Nanotechnol.} \textbf{4} 510-6

\bibitem{Jensen_NL_2004} Jensen L E, Bj\"ork M T, Jeppesen S, Persson A I, Ohlsson B J and Samuelson L 2004 Role of surface diffusion in chemical beam epitaxy of InAs nanowires \textit{Nano Lett.} \textbf{4} 1961-4

\bibitem{Lehmann_NL_2013} Lehmann S, Wallentin J, Jacobsson D, Deppert K and Dick K A 2013 A general approach for sharp crystal phase switching in InAs, GaAs, InP, and GaP nanowires using only group V flow \textit{Nano Lett.} \textbf{13} 4099-105

\bibitem{Gluschke_NL_2018} Gluschke J G, Seidl J, Lyttleton R W, Carrad D J, Cochrane J W, Lehmann S, Samuelson L and Micolich A P 2018 Using ultrathin parylene films as an organic gate insulator in nanowire field-effect transistors \textit{Nano Lett.} \textbf{18} 4431-9

\bibitem{Suyatin_NT_2007} Suyatin D B, Thelander C, Bj\"ork M T, Maximov I and Samuelson L 2007 Sulfur passivation for ohmic contact formation to InAs nanowires \textit{Nanotechnology} \textbf{18} 105307

\bibitem{Ji_ME_2008} Ji F, Xu J P, Lai P T and Guan J G 2008 A fringing-capacitance model for deep-submicron MOSFET with high-k gate dielectric \textit{Microelectron. Reliab.} \textbf{48} 693-7

\bibitem{Song_IEEE_2006} Song J Y, Choi W Y, Park J H, Lee J D and Park B-G 2006 Design optimization of gate-all-around (GAA) MOSFETs \textit{IEEE Trans. Nanotechnol.} \textbf{5} 186-91

\bibitem{Zou_IEEE_2011} Zou J, Xu Q, Luo J, Wang R, Huang R and Wang Y 2011 Predictive 3-D Modeling of Parasitic gate capacitance in gate-all-around cylindrical silicon nanowire MOSFETs \textit{IEEE Trans. Electron Devices} \textbf{58} 3379-87

\bibitem{Gupta_JS_2013} Gupta S K and Baishya S 2013 Modeling of cylindrical surrounding gate MOSFETs including the fringing field effects \textit{J. Semicond.} \textbf{34} 074001

\bibitem{Lee_SSE_2007} Lee C-W, Yun S-R-N, Yu C-G, Park J-T and Colinge J-P 2007 Device design guidelines for nano-scale MuGFETs \textit{Solid-State Electron.} \textbf{51} 505-10

\bibitem{Heedt_NS_2015} Heedt S, Otto I, Sladek K, Hardtdegen H, Schubert J, Demarina N, L\"uth H, Gr\"utzmacher D and Sch\"apers T 2015 Resolving ambiguities in nanowire field-effect transistor characterization \textit{Nanoscale} \textbf{7} 18188-97

\bibitem{Cord_NSPMP_2009} Cord B, Yang J, Duan H, Joy D C, Klingfus J and Berggren K K 2009 Limiting factors in sub-10~nm scanning-electron-beam lithography \textit{J. Vac. Sci. Technol. B} \textbf{27} 2616-21

\bibitem{Mohammad_book_2010} Mohammad M A 2010 The interdependence of exposure and development conditions when optimizing low-energy EBL for nano-scale resolution \textit{Lithography} ed Taras Fito (Rijeka: IntechOpen) Ch. 16

\end{thebibliography}
\end{document}